\documentclass[review]{elsarticle}

\usepackage{lineno,hyperref}
\modulolinenumbers[5]

\usepackage[caption=false,font=footnotesize]{subfig}
\usepackage{mathtools} 
\usepackage{algpseudocode} 
\usepackage{algorithm} 
\algnewcommand{\LineComment}[1]{\State \(\triangleright\) #1}  
\usepackage{amsmath}
\usepackage{amssymb}
\usepackage{multirow} 
\usepackage{array} 
\newcolumntype{C}[1]{>{\centering}m{#1}}
\newcolumntype{C}[1]{>{\centering\let\newline\\\arraybackslash\hspace{0pt}}m{#1}}
\usepackage{float}
\usepackage{graphicx}
\usepackage{xspace}
\usepackage{listings}     
\usepackage{xcolor}
\DeclareMathOperator*{\maxi}{maximize}
\def \B {\ensuremath{\mathcal{B}}\xspace}
\def \U {\ensuremath{\mathcal{U}}\xspace}
\def \X {\ensuremath{\mathcal{X}}\xspace}

\newcommand{\Reals}[0]{\ensuremath{\mathbb{R}}}

\journal{Journal of \LaTeX\ Templates}

\begin{document}

\begin{frontmatter}

\title{Solving Batched Linear Programs on GPU and Multicore CPU}

\author{Amit Gurung*}
\ead{amitgurung@nitm.ac.in}

\author{Rajarshi Ray}
\ead{rajarshi.ray@nitm.ac.in}

\cortext[mycorrauthor]{Corresponding author}

\address{Department of Computer Science \& Engineering,	National Institute of Technology Meghalaya, Shillong - 793003, India}

	\begin{abstract}
Linear Programs (LPs) appear in a large number of applications and offloading them to the GPU is viable to gain performance. Existing work on offloading and solving an LP on GPU suggests that performance is gained from large sized LPs (typically 500 constraints, 500 variables and above). In order to gain performance from GPU for applications involving small to medium sized LPs, we propose batched solving of a large number of LPs in parallel. In this paper, we present the design and CUDA implementation of our batched LP solver library, keeping memory coalescent access, reduced CPU-GPU memory transfer latency and load balancing as the goals. The performance of the batched LP solver is compared against sequential solving in the CPU using an open source solver GLPK (GNU Linear Programming Kit). The performance is evaluated for three types of LPs. The first type is with the initial basic solution as feasible, the second type is with the initial basic solution as infeasible and the third type is with the feasible region as a Hyperbox. For the first type, we show a maximum speedup of $18.3\times$ when running a batch of $50k$ LPs of size $100$ ($100$ variables, $100$ constraints). For the second type, a maximum speedup of $12\times$ is obtained with a batch of $10k$ LPs of size $200$. For the third type, we show a significant speedup of $63\times$ in solving a batch of nearly $4$ million LPs of size 5 and $34\times$ in solving 6 million LPs of size $28$. In addition, we show that the open source library for solving linear programs-GLPK, can be easily extended to solve many LPs in parallel with multi-threading. The thread parallel GLPK implementation runs $9.6\times$ faster in solving a batch of $1e5$ LPs of size $100$, on a $12$-core Intel Xeon processor. We demonstrate the application of our batched LP solver in the domain of state-space exploration of mathematical models of control systems design.
	
	\end{abstract}


\end{frontmatter}

\section{Introduction\label{sec:Introduction}}

Computations which were traditionally purely carried out in the CPU are increasingly being computed with CPU and GPU in heterogeneity by offloaded expensive data parallel tasks to a GPU for accelerating performance.  Some of the application domains where GPU has been used to accelerate performance include medical image processing \cite{birk2014gpu,Yan2008}, weather research and forecasting (WRF) \cite{Michalakes2008}, Proteomics (to speed-up peptide spectrum matching \cite{Engineering2011}), signal processing for radio astronomy\cite{clark2012accelerating}, simulation of various physical and mechanical systems (using variants of Monte Carlo algorithm)\cite{karimi2011high,lim2015high} and large scale graph processing \cite{Shirahata2012}. However, gaining performance from a GPU requires insights on its architecture in order to have an effective load balancing, efficient memory access and an effective mapping of computations in the SIMD paradigm of computing.

Linear Programming is a method of maximizing or minimizing a linear objective function subject to a set of linear constraints. Linear programs (LPs) appear extensively in a large number of application domains such as business process modeling to maximize profit, economics to design optimized demand-supply model (for example Leontief Input-Output model \cite{chandra1988economic}), optimal cost and transport assignment in transportation problem \cite{munkres1957algorithms}, optimal job scheduling \cite{drozdowski2006scheduling} and optimize packets routing in computer networks, to name just some. 

In this work, our focus is on CPU-GPU heterogeneous computations that, in particular, requires solving a large number of LPs. Our work is on the setting that computations begin in a CPU where LPs are created and then offloaded to a GPU for an accelerated solution. The solutions are transferred back to the CPU from the GPU for further processing. There has been prior work in this direction with parallel implementation of algorithms to solve LPs on a GPU, like the simplex and revised simplex algorithm \cite{lalami2011multi,bieling2010efficient,DBLP:conf/ipps/SpampinatoE09}. However, the performance gain is reported only when offloading large LPs of size $500$ (500 constraints, 500 variables) and above. Prior works state that for small size LPs, the time spent in offloading the LPs from CPU to GPU memory is more than the time gained with parallel solution in the GPU. Therefore, how can applications requiring to solve small to medium size LPs exploit the power of a GPU, remains a research challenge.   

Our work in this paper target application that involves solving small to medium size LPs, but many of them. The existing work of offloading LPs to GPU does not provide acceleration in such applications due to small-medium size LPs. We therefore propose to use GPUs to solve not a single LP at a time, but to batch them and solve them simultaneously. We show that with batched computation, the performance gain with parallelism is more than the performance loss in transferring LP tasks from CPU-GPU memory, even for small size LPs (e.g. LPs of size 5). We present a CUDA C/C++ implementation of our library which implements the simplex method \cite{DantzigT97}, with an effort to keep coalescent memory accesses, efficient CPU-GPU memory transfer and effective load balancing. To the best of our knowledge, this is the first work in the direction of batched LP solving in the GPU. Batched computations in GPU to draw performance is, however emerging as a technique in general \cite{abdelfattah2016performance,jhurani2015gemm}. The library source and necessary instructions for repeatability evaluation can be found at \url{https://bitbucket.org/rajgurung777/simplexprojects}.

We report solutions of LPs of dimension up to $511 \times 511$ ($511$ variables, $511$ constraints) with our library. We show that beyond a sufficiently large batch size, our implementation shows significant gain in performance compared to solving them sequentially in the CPU using the GLPK library \cite{GLPK}, an open source LP solver. We also report our observations on two pivot selection rules in the simplex method implemented in the library. In addition, we present a technique to solve a special class of LP when the feasible region is a hyper-rectangle and show that these can be solved cheaply without using the simplex algorithm. We implement this special case LP solver as part of the library.

Finally, we attempt to address the problem that GLPK implementation is not \emph{thread safe}. By \emph{not thread safe}, we mean that multiple threads running local instances of the GLPK object is not safe. As a solution, we show the necessary changes to make it thread safe and report performance gain with multi-threaded solving of many LPs in a multi-core architecture.

The rest of the paper is organized as follows. Related works are discussed in Section \ref{sec:Related-Work}. In Section \ref{sec:LinearProgramming}, we discuss the simplex method that is needed to appreciate the rest of the paper. Section \ref{sub:Parallel-Algorithm} illustrates our CUDA implementation for solving batched LPs on GPU, with memory coalescence, effective load balancing and efficient GPU-CPU memory transfer using CUDA streams to gain performance. In Section \ref{sec:parallelGLPK}, we present the implementation and experimental results of a thread safe GLPK for solving multiple LPs using multi-threading. Section \ref{sec:HyperboxLP} shows the performance of our CUDA implementation for solving a special class of LP problems in batches in comparison to solving the same LPs sequentially with GLPK. In Section \ref{sec:HybridSystem}, we show an application of our batched LP solver GPU library in the domain of model based analysis of control systems design.

\section{Related Work \label{sec:Related-Work}}

A multi-GPU implementation of the simplex algorithm in \cite{lalami2011multi} reports a speedup of $2.93\times$ on LP problems of dimension $1000 \times 1000$. An average speedup of $12.7\times$ has been reported for the larger problems of dimension $8000 \times 8000$ or higher on a single GPU. An implementation of the revised simplex method using inbuilt graphics library (OpenGL) is reported in \cite{bieling2010efficient}. An average speedup of $18\times$ has been reported, compared to the GLPK library, for problems of size $600 \times 600$ or higher. A GPU implementation of the revised simplex algorithm is also reported in \cite{DBLP:conf/ipps/SpampinatoE09} with a speedup of $2\times$ to $2.5\times$ in comparison to a serial ATLAS-based CPU implementation for LPs of dimension $1400$ up to $2000$. Automatically Tuned Linear Algebra Software (ATLAS\cite{whaley2011atlas}) is a software library for linear algebra providing an implementation of the BLAS (Basic Linear Algebra Subprograms) APIs for C and Fortran. BLAS\cite{BLAS-Jack} is a specification that prescribes routines for basic vector and matrix operations. BLAS implementation is optimized for performance on a specific architecture. We observed that almost all the works report speedup only for large size LP problems (typically of dimension $500 \times 500$ or above) compared to the sequential CPU implementations.

\section{Linear Programming \label{sec:LinearProgramming}}

A linear program in \emph{standard form} is maximizing an objective function under the given set of linear constraints, represented as follows:

\begin{equation}
maximize\qquad\sum_{j=1}^{n}c_{j}x_{j}\label{eq:Objective_Function}
\end{equation}

subject to the constraints

\begin{equation}
\sum_{j=1}^{n}a_{ij}x_{j}\le b_{i}\quad for\quad i=1,2,...,m\label{eq:constraints}
\end{equation}

and
\begin{equation}
x_{j}\ge0\quad for\quad j=1,2,...,n\label{eq:NonNegativeConstraints}
\end{equation}
In Expression (\ref{eq:Objective_Function}) , $\sum_{j=1}^{n}c_{j}x_{j}$ is the objective function to be maximized and Inequality (\ref{eq:constraints}) shows the $m$ constraints over $n$ variables. Inequality (\ref{eq:NonNegativeConstraints}) shows the non-negativity constraints over $n$ variables. An LP in \emph{standard form} can be converted into \emph{slack form} by introducing $m$ additional \textbf{slack variables} ($x_{n+i}$), one for each inequality constraint, to convert it into an equality constraint, as shown below:

\begin{equation}
x_{n+i}=b_{i}-\sum_{j=1}^{n}a_{ij}x_{j},\:for\:i=1,...,m\label{eq:constraints-1}
\end{equation}
An algorithm that solves LP problems efficiently in practice is the \emph{simplex method} described in \cite{DantzigT97}. The variables on the left-hand side of the Equation (\ref{eq:constraints-1}) are referred to as \textbf{basic variables} and those on the right-hand side are \textbf{non-basic variables}. The \emph{initial basic solution} of an LP is obtained by assigning its non-basic variables to 0. The \emph{initial basic solution} may not be always feasible (when one or more of the $b_i$s are negative, resulting in the violation of the non-negativity constraint). For such LPs, the simplex method employs a two-phase algorithm. A new \textbf{auxiliary LP} is formed by having a new objective function $z$, which is the sum of the newly introduced \textbf{artificial variables.} The \textbf{simplex algorithm} is employed on this auxiliary LP and it is checked if the optimal solution to the objective function is $0$. If a zero optimal is found, then the original LP has feasible solution and the simplex method initiates for Phase II. Therefore, LPs with infeasible initial basic solution takes more time to be solved.

\subsection{The Simplex Algorithm \label{sub:The-Simplex-Algorithm}}

The simplex algorithm is an iterative process of solving an LP problem. Each iteration of the simplex algorithm attempts to increase the value of the objective function by replacing one of the basic variables (also known as the \textbf{leaving variable}), by a non-basic variable (called the \textbf{entering variable}). The exchange of these two variables is obtained by a \emph{pivot operation}. The index of the leaving and the entering variables are called the pivot row and pivot column respectively. The simplex algorithm iterates on a tabular representation of the LP, called the \textbf{simplex tableau}. The simplex tableau stores the coefficients of the non-basic, slack and artificial variables in its rows. It contains auxiliary columns for storing intermediate computations. For our discussion, we consider a tableau of size $p \times q$, where $p=m+1$ and $q=n+\textit{sum of slack and artificial variables} + 2$. The ($m+1$) the row stores, the best solution to the objective function found so far, along with the coefficients of the non-basic variables in the objective function.

\begin{figure}[h]
\centering{}
   \includegraphics[scale=0.9]{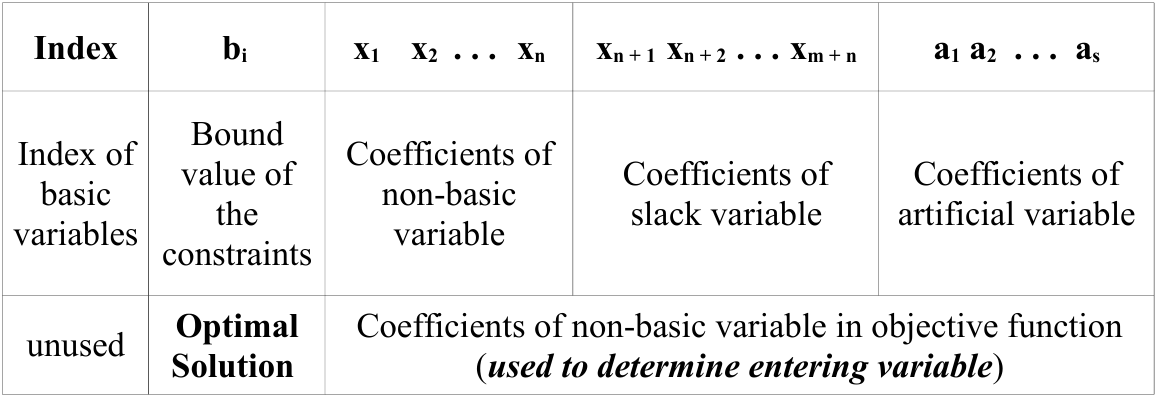}
\caption{Formation of the Simplex Tableau.\label{fig:SimplexTableau}}
\end{figure}

There are two auxiliary columns, the first column stores the index of the basic variables and the second stores $b_i$'s of inequality (\ref{eq:constraints}). A schematic of the simplex tableau is shown in Figure \ref{fig:SimplexTableau}.

\textbf{Step 1: Determine the entering variable. }

At each iteration, the algorithm identifies a new \emph{entering variable} from the non-basic variables. It is called an entering variable since it enters the set of basic variables. The choice of the entering variable is with the goal that increasing its value from 0 increases the objective function value. The index of the entering variable is referred to as the \emph{pivot column}. The most common rule for selecting an entering variable is by choosing the index $e$ of the maximum in the last row of the simplex tableau (excluding the current optimal solution). 

\textbf{Step 2: Determine the leaving variable. }

Once the pivot column is determined (say $e$), the algorithm identifies the row index with the minimum positive ratio $(b_i/-a_{e,i})$, say $\ell$, called the \emph{pivot row}. The variable $x_\ell$ is called the leaving variable because it leaves the set of basic variables. This ratio represents the extend to which the entering variable $x_e$ (in step 1) can be increased without violating the constraints. 

\textbf{Step 3: Obtain the new improved value of the objective function. }

The algorithm then performs the \emph{pivot operation} which updates the simplex tableau such that the new set of basic variables are expressed as a linear combination of the non-basic ones, using substitution and rewriting. An improved value for the objective function is found after the pivot operation.

The above steps are iterated until the halt condition is reached. The halt condition is met when either the LP is found to be \emph{unbounded} or the \textbf{optimal solution} is found. An LP is unbounded when no new leaving variable can be computed, i.e. when the ratio ($b_i/ - a_{e,i}$) in step 2 is either negative or undefined for all $i$. An optimal solution is obtained when no new entering variable can be found, i.e., the coefficients of the non-basic variables in the last row of the tableau are all negative values

\section{Simultaneous Solving of Batched LPs on GPU \label{sub:Parallel-Algorithm}}
We present our CUDA implementation that solves fixed size batched LPs in parallel on a GPU. In this discussion, we shall refer a CPU by \emph{host} and a GPU by \emph{device}.

\subsection{CPU-GPU Memory Transfer and Load balancing}

First, we allocate device memory (global memory) from the host, that is required for creating a simplex tableau for every LP in the batch. The maximum number of LPs that can be batched depends on the size of the device global memory. The tableau for every every LP in the batch is populated with all the coefficients and indices of the variables in the host side, before transferring to the device. To speedup populating the tableau in the host, we initialize the tableau in parallel using OpenMP threads. Once initialized, the simplex tableaux are copied from the host to the device memory (referred to as H2D-ST in Figure \ref{fig:DataOverlapWithKernel}). The GPU kernel modifies the tableaux to obtain solution using the simplex method and the results for every LP in the batch is copied back from the device to the host memory (referred to as D2H-res in Figure \ref{fig:DataOverlapWithKernel}). We discuss further on our CPU-GPU memory transfer using CUDA streams for efficiency in Section \ref{sub:streams}.

\paragraph{Load Balancing}

\begin{figure}[h]
\centering{}
   \includegraphics[scale=0.35]{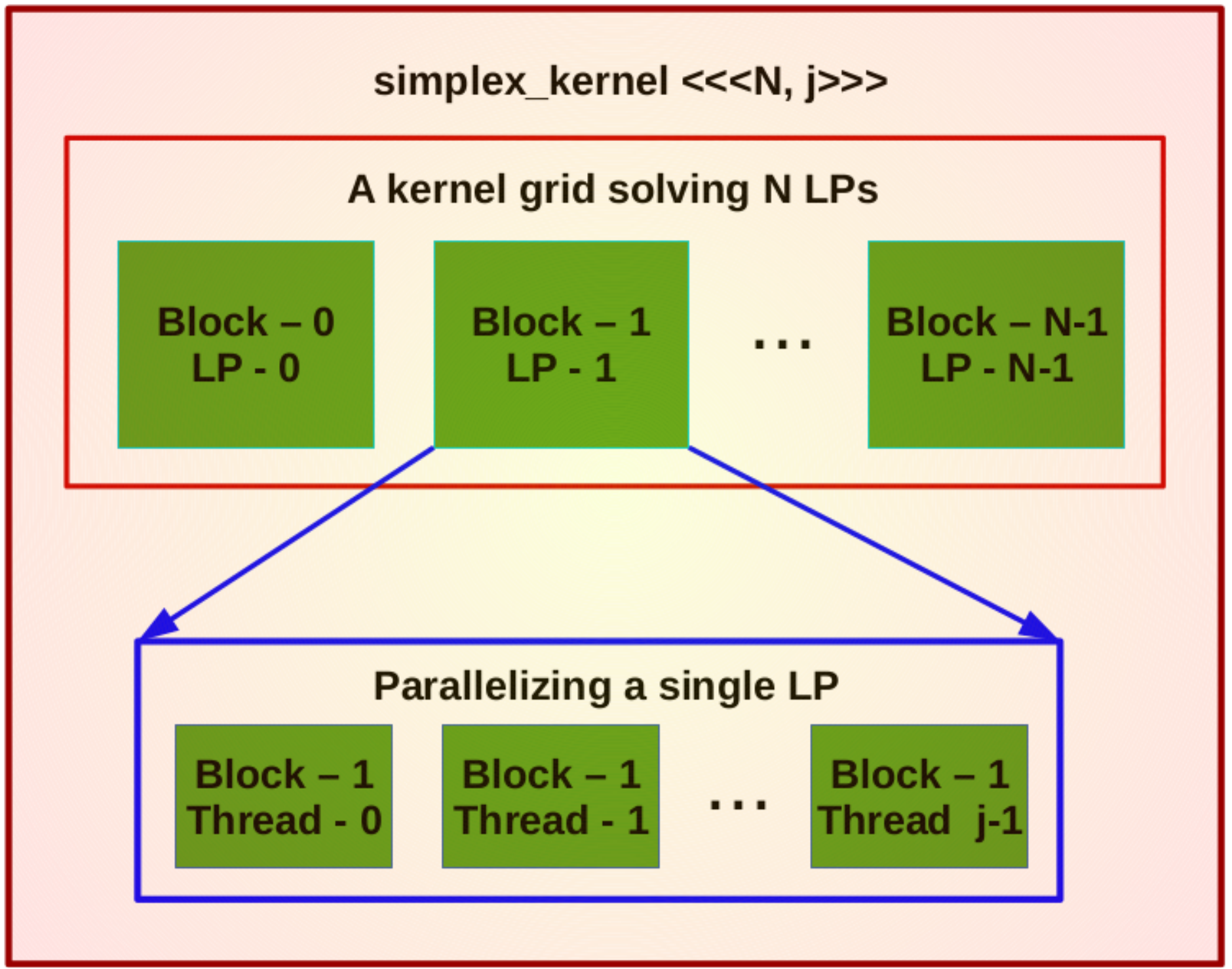}
\caption{Visualization of how threads are mapped to solve $N$ LPs in GPU. Each block is mapped to an LP and $j$ threads are assigned to parallelize a single LP.\label{fig:SimplexKernel}}
\end{figure}

We assign a CUDA block of threads to solve an LP in the batch. Since blocks are scheduled to Streaming Multiprocessors (SMs), this ensures that all SMs are busy when there are sufficiently large number of LPs to be solved in the batch. As CUDA blocks execute asynchronously, such a task division emulates solving many LPs independently in parallel. Moreover, each block is made to consist of $j$ ($\ge q$) threads, which is a multiple of $32$, as threads in GPU are scheduled and executed as warps. The block of threads is utilized in manipulating the simplex tableau in parallel, introducing another level of parallelism. In Figure \ref{fig:SimplexKernel}, we show a block diagram of our parallel implementation on the GPU.  

.
\subsection{Simplex Algorithm Implementation}

Finding the pivot column in \textbf{step 1} of the simplex algorithm above requires to determine the index of the maximum value from the last row of the tableau. We have parallelized \textbf{step 1} by utilizing $n$ (out of $j$) threads in parallel to determine the pivot column using \textbf{parallel reduction} described in \cite{Harris2007}. A parallel reduction is a technique applied to achieve data parallelism in GPU when a single result (e.g. min, max) is to be computed out of a large number of data. We have implemented a parallel reduction by using two auxiliary arrays, one for storing the data and the other for storing the indices of the corresponding data. The result of a parallel reduction algorithm provides us the maximum value in the first array and its corresponding index in the other array. 

We also applied parallel reduction in \textbf{step 2} by utilizing $m$ (out of $j$) threads in parallel to determine the pivot row ($m$ being the row-size of the simplex tableau). Using parallel reduction in step 2 requires other modifications. It involves finding a minimum positive value from a vector of ratios (as described in Step 2 above) and therefore ratios which are not positive needs to be excluded from the minimum computation. This leads to a conditional statement in the parallel reduction algorithm and degrades performance due to warp divergence. Even if we re-size the vector to store only the positive values, the kernel will still require conditional statements to check the thread IDs that need to process this smaller size vector. To overcome performance degradation with conditional statements, we substituted a large positive number in place of ratios that are negative or undefined. This creates a vector that is suitable for parallel reduction in our kernel implementation.

Data parallelism is also employed in the pivot operation in \textbf{step 3}, involving substitution and re-writing, using the ($m-1$) threads (out of $j$ threads on the block).

There are a number of pivot selection rules that could be applied in step 1. In this work, we have experimented with two pivot selection rules, to study its effect on the performance of simplex algorithm in the GPU. We describe these pivot selection rules below:

\paragraph{Largest Positive Coefficient (LPC):} We take the index of the maximum positive coefficient in the last row of the simplex tableau (step 1 in the above algorithm).
\paragraph{Random Positive Coefficient (RPC):} Instead of choosing the index of the maximum positive coefficient as in LPC, we choose a random index having a positive coefficient from the last row of the simplex tableau. Although this rule is generally not efficient since it may result in more iterations in the algorithm, our purpose is to see its effect in the context of a GPU implementation since it requires no overhead of parallel reduction unlike the LPC rule. The choice of this pivot selection rule may be appropriate in the context of the GPU as a warp of $32$ threads can read simultaneously $32$ values in only one cycle and can assign only the index containing the positive value to a shared variable, as the pivot column. Therefore, RPC rule seems to incur less overhead in simplex iterations in GPU compared to the LPC rule. However, it remains to be experimented if this gain dominates the loss of performance due to the possible extra iterations.
Our observations on the performance using the above mentioned pivot selection rules are illustrated in Section. \ref{sec:AnalyseMultiLPs-GPU}.

\subsection{Memory Coalescent Access \label{sub:memoryCoalescing}}

In this section, we discuss our efforts of keeping a coalescent access to global memory to reduce performance loss due to cache misses. When all threads of a warp access contiguous region of the memory, it is coalesced and this ensures improved performance due to high cache hit rate. However, if the access to memory is not coalesced, then the memory controller undergoes cache block replacements that incur delays and degrades the performance in GPU. 

As discussed earlier, we use global memory to store the simplex tableaux of the LPs in a batch as described in Section \ref{sub:The-Simplex-Algorithm} ( Global memory being the largest can accommodate many tableaux). We store the simplex tableau in memory as a two-dimensional array. High level languages like C and C++ uses the row-major order by default for representing a 2-dimensional array in the memory. CUDA is an extension to C/C++ and also uses the row-major order The choice of row or column major order representation of two-dimensional arrays plays an important role in deciding the efficiency of the implementation, depending on whether the threads in a warp access the adjacent rows or adjacent columns of the array and what is the offset between the consecutive rows and columns.

We use the term \textit{column-operation}, when element of all rows from a specific column is accessed simultaneously by each thread in a warp. If the array is in a row-major order, then this operation is not a coalesced memory access, as each thread accesses elements from the memory separated by the size equal to the column-width of the two dimensional array. When elements of a specific row are accessed simultaneously by each thread of a warp, we called this a \textit{row-operation}. Note that for a two dimensional array stored in row-major order, a row-operation is coalesced since each thread accesses data from contiguous region in the memory.

We show below that in the simplex algorithm described above, there are more column-operations than row operations and thus, storing our data (i.e. simplex tableau) in a column-major order would ensure higher coalesced memory accesses.  

\begin{figure}[h]
\centering{}
   \includegraphics[width=8.6cm,height=6cm]{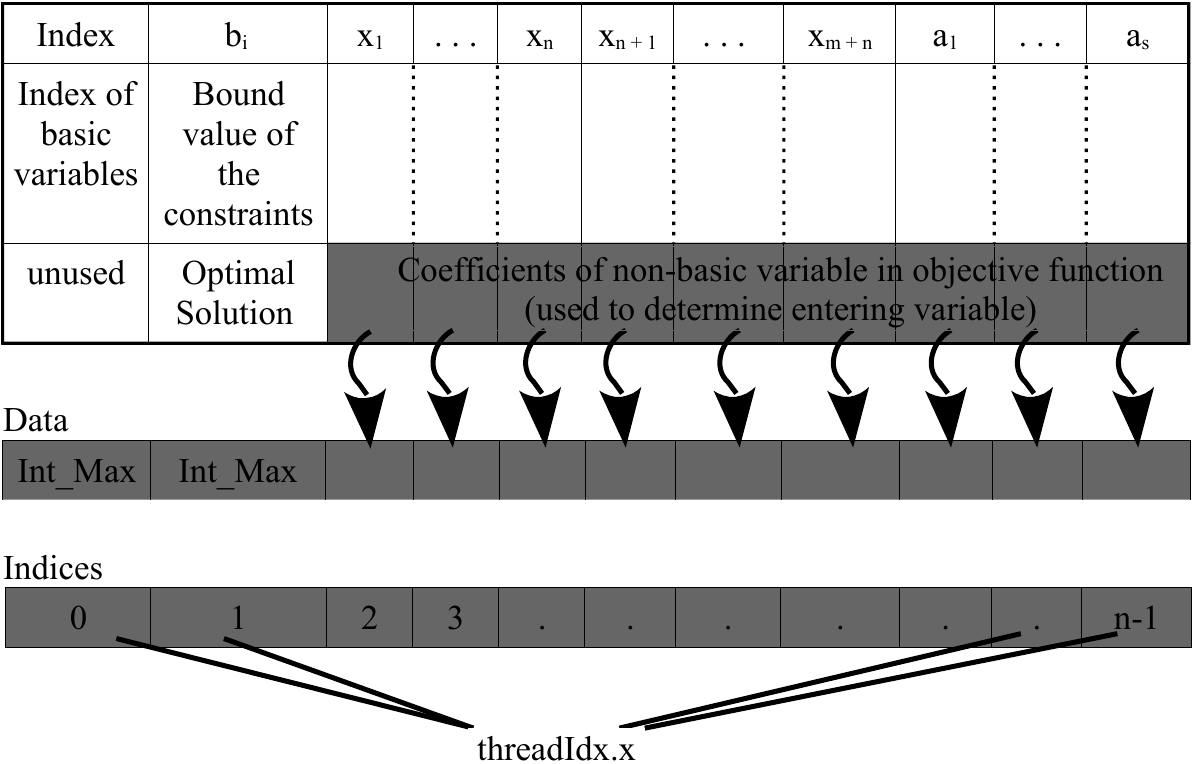}
\caption{Showing the Simplex Tableau along with two separate arrays, \textit{Data} to store the coefficients of the objective function and \textit{Indices} to keep track of the indices of the corresponding values in the \textit{Data} array.\label{fig:Step1}}
\end{figure}

\textbf{Step 1} of the simplex algorithm determines the entering variable (also known as the pivot column), which requires finding the index of the maximum positive coefficient (in case of LPC rule) from the last row. This requires a row-operation and as mentioned in Section \ref{sub:Parallel-Algorithm}, we use parallel reduction using two auxiliary arrays, \textit{Data} and \textit{Indices} as shown in Figure \ref{fig:Step1}. Although accessing from the last row of the simplex tableau is not coalesced (due to our column-major ordering) but copying into the Data (and Indices) array is coalesced and so is the parallel reduction algorithm on the Data (and Indices) array. We used the technique of \emph{Parallel Reduction: Sequential Addressing} in \cite{Harris2007}, a technique that ensures coalesced memory access.

\begin{figure}[h]
\centering{}
   \includegraphics[scale=0.7]{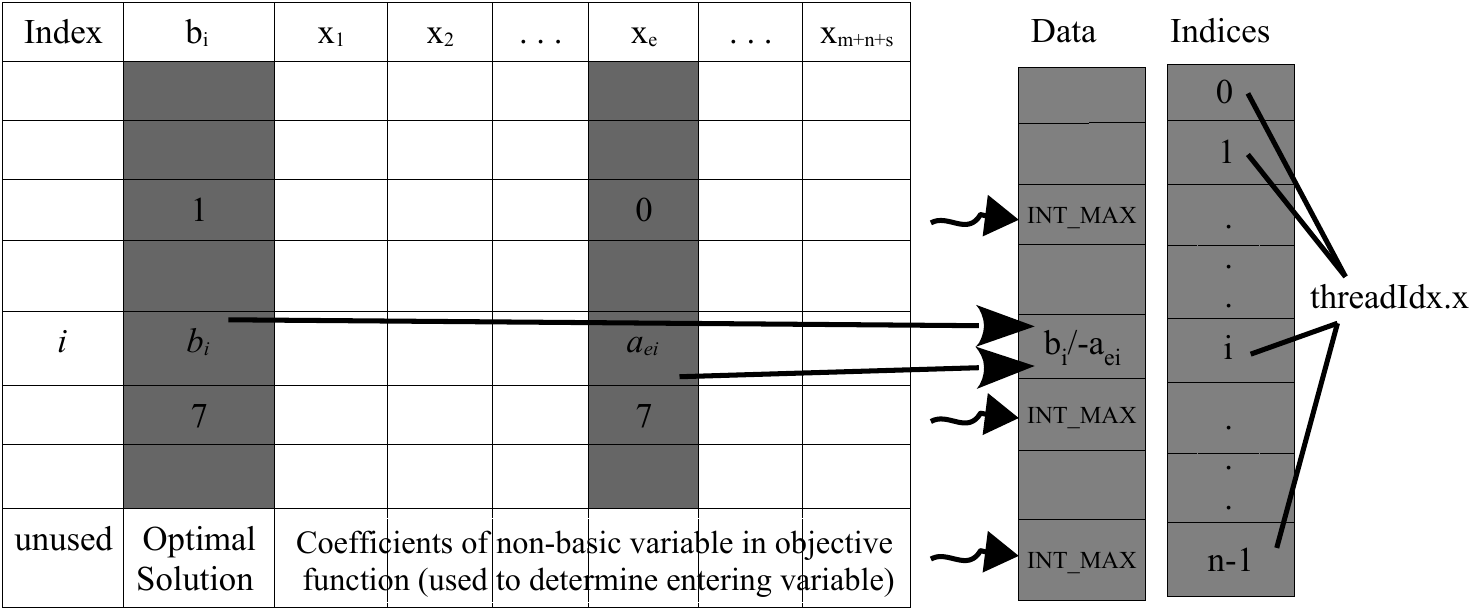}
\caption{Showing the Simplex Tableau along with two separate arrays, \textit{Data} to store the positive ratio and \textit{Indices} to keep track of the indices of the corresponding values in the \textit{Data} array. Ratios that reduces to negative or undefined are replaced by a large value denoted by INT\_MAX. \label{fig:Step2}}
\end{figure}

\textbf{Step 2} of the simplex algorithm determines the leaving variable (also called the pivot row) by computing the row index with the minimum positive ratio $(b_i/-a_{e,i})$, as described in Section \ref{sub:The-Simplex-Algorithm}. This requires two column-operations involving the access to all elements from columns $b_i$ and $a_{e,i}$ as shown in Figure \ref{fig:Step2}. To compute the row index with the minimum positive ratio, we use parallel reduction as described above in Section \ref{sub:Parallel-Algorithm}. Our tableau being stored in a column-major order, access to columns $b_i$ and $a_{e,i}$ are both coalesced. The ratio and its corresponding indices (represented by the thread ID) are stored in the auxiliary arrays, \textit{Data} and \textit{Indices} which is also coalesced. Like in Step 1, we use the same technique of \emph{Parallel Reduction: Sequential Addressing} in \cite{Harris2007} for coalesced memory access.

\textbf{Step 3} performs the pivot operation that updates the elements of the simplex tableau and is the most expensive of the three steps. It first involves a non-coalesce row-operation which computes the new modified pivot row (denoted by the index $\ell$) as \{$NewPivotRow_{\ell} = OldPivotRow_{\ell} \div PE$\}, where PE is the element in cell in the intersection of the pivot row and the pivot column for that iteration, known as the pivot element. The modified row ($NewPivotRow_{\ell}$) is then substituted to update each element of all the rows of the simplex tableau, using the formula $NewRow_{ij} = OldRow_{ij} - PivotCol_{ie} * NewPivotRow_{\ell j}$ (see the code Listing \ref{code:step3} below). The elements of the pivot column are first stored in an array  named $PivotCol$ which is a column-operation, and so is coalesced, due to the column-major representation of the tableau. The crucial operation is updating each $j^{th}$ element for every $i^{th}$ row (except the pivot row $\ell$) of the simplex tableau, which requires a nested for-loop operation. We have parallelized the outer for-loop that maps the rows of the simplex tableau. Our data being represented in a column-major order, so parallel access to all rows for each element in the $j^{th}$ column of the inner for-loop is coalesced.

\begin{lstlisting}[language=c, caption={Showing code fragment for step 3 that updates the simplex tableau.}, label={code:step3}]

for (int i=0;i<rows;i++) { //Parallelized outer loop to map each i with the thread ID
  for (int j=0;j<cols;j++) {
    NewRow[i][j] = OldRow[i][j] - PivotCol[i] * NewPivotRow[l][j]; //l index of pivot row
  }
}
\end{lstlisting}

\begin{figure*}[!htb]
\centering{}
   \includegraphics[scale=0.65]{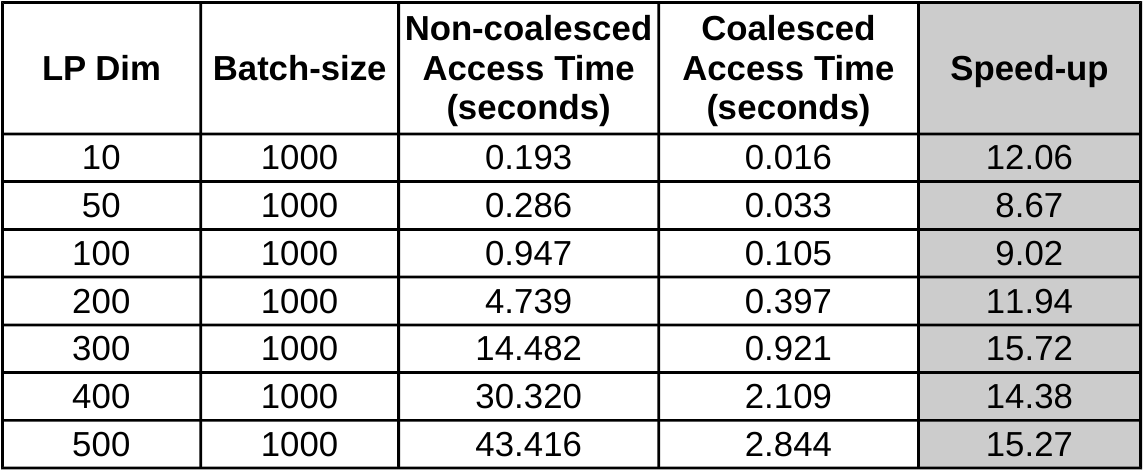}
\caption{Showing the time taken to solve batched LP due to coalesced and non-coalesced memory access on GPU, in LPC implementation for LPs with initial basic solution as feasible. \label{fig:NonCoalescedResult}}
\end{figure*}

To verify the performance gained due to coalesced memory access, we have experimented with \textbf{Step 3} which is the most expensive of all steps in the simplex algorithm, by modifying it to have non-coalesced memory access. In the code Listing \ref{code:step3}, we interchange the inner for-loop with the outer loop (loop interchange, a common technique to improve cache performance\cite{hennessy2011computer}). This loop interchanging converts the Step 3 to have non-coalesced memory access since our simplex tableau is represented in a column-major order. Figure \ref{fig:NonCoalescedResult} presents the experimental results to show the gain in performance when the access to memory is coalesced as compared to non-coalesced access. Clearly, the result has shown a significant gain in performance on a Tesla K40c card, implementing the LPC pivot selection rule for LPs with initial basic solution as feasible.

We observed that step 1 has a row-operation, step 2 has two column-operations and step 3 has a row and a column operation each along with a nested for-loop which can be expressed both row as well as column operations. Clearly, there are more column-operations than rows. However, the size of column is more than row of our simplex tableau, therefore, one can experiment on the row-major layout of the tableau, to determine if this representation has higher coalesced memory accesses.

\subsection{Overlapping data transfer with kernel operations using CUDA Streams \label{sub:streams}}

\begin{figure*}[!htb]
\centering{}
   \includegraphics[scale=0.65]{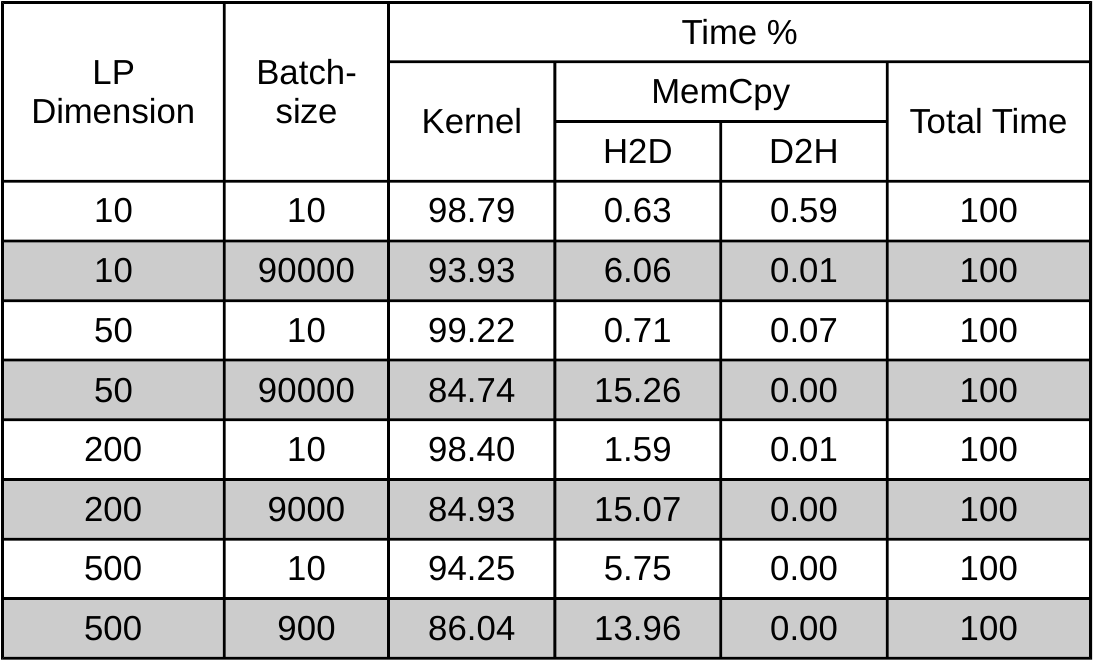}
\caption{Showing the profile report obtained using \textbf{nvprof} tool in our LPC implementation for LPs with an initial basic solution as feasible. H2D - stands for host to device and D2H indicates device to host memory copies respectively. \label{fig:profileResult}}
\end{figure*}

The memory bandwidth of host-device data copy is a major bottleneck in CUDA applications. We use Nvidia's profiling tool \textbf{nvprof} \cite{CudaGuide} to profile time for memory transfer and kernel operation for our implementation discussed above in Section \ref{sub:Parallel-Algorithm}. The result of profiling in a Tesla K40c card, implementing the LPC pivot selection rule for LPs with an initial basic solution as feasible, is reported in Figure \ref{fig:profileResult}. We observed that, for a small batch-size problem (e.g. $10$ in the Figure \ref{fig:profileResult}), the memory copy operation is a maximum of $5.75\%$, where as for bigger batch-size problem the memory copy operation is in the range of $10-15\%$ and above. Although, the value is not substantial for significant performance tuning, but it cannot be ignored either. 

A standard technique to improve performance in CUDA applications is by using CUDA streams which allow overlapping memory copies with kernel execution. A stream in CUDA consists of a sequence of operations, which is executed on the device in the order in which they are issued by the host procedure. These different sequence of operations not only can be interleaved, but can also be executed concurrently in order to gain higher performance as described in \cite{OverlapHarris2012}.

\begin{figure*}[!htb]
\centering{}
   \includegraphics[scale=0.45]{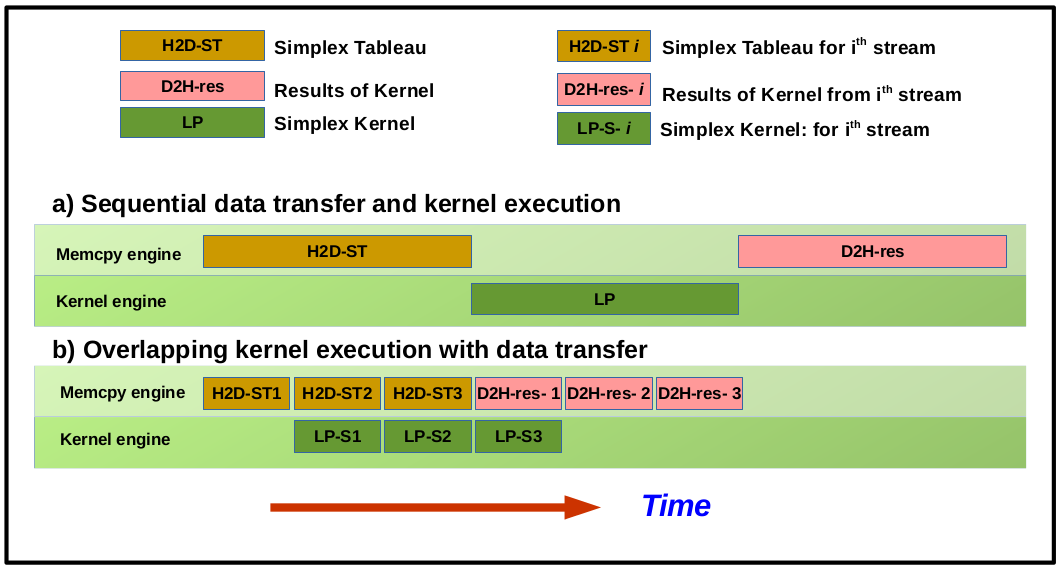}
\caption{ Showing the gain in time due to overlapping kernel execution with data transfer compared to sequential data transfer and kernel execution. The time required for host-to-device(H2D), device-to-host(D2H) and kernel execution are assumed to be same.\label{fig:DataOverlapWithKernel}}
\end{figure*}

A GPU in general, has a separate kernel and a copy engine. All kernel operations are executed using the kernel engine and memory copy operations to and from the device is performed by the same copy engine. However, some GPU cards have two copy engines, one each for copying data to and from the device, to achieve higher performance on the GPU. Figure \ref{fig:DataOverlapWithKernel}, illustrates the overlapping of kernel executions with data copy, when the GPU has only one kernel and copy engine each. To obtain maximum performance in such GPU configuration, streaming by batching similar operations provides more overlap of copies with kernel executions. This is done by adding all host-to-device copy to the different streams followed by all kernel launches and device-to-host data copies. When there is to copy engines, looping the operations in the order of a host-to-device copy followed by kernel launch and device-to-host copy, for all streams would yield higher performance than the former method. However, for all devices with compute capability 3.5 and above, both the methods yield same performance, due to the Hyper-Q \cite{HyperQEg} feature enabled in them.

Higher number of CUDA streams achieves higher concurrency and interleaving among operations, but it involves stream creation overhead. The number of CUDA streams that gives optimal performance is found by experimentation. From our experimental observations, we conclude that with varying batch size and LP dimension to be solved, the optimal number of streams also varies. In this paper, we have reported the results with $10$ streams for batch size higher than $100$ LPs and only $1$ stream when the batch size is less than $100$ (for LPs of any dimension).

\subsection{Limitations of the Implementation \label{sub:implementation}}
The memory required for an LP (i.e., a tableau) in our implementation can be approximately computed as:

\begin{equation}
Y=\{(m+1) \times cols \times dataSize +x\}
\label{eq:SingleLPSize}
\end{equation}

$\indent \indent  \indent  \indent  \indent  \indent cols = (var+slack+arti+2)$ \\
$\indent \indent \indent \indent  \indent \indent  \indent dataSize = sizeof(DType)$ \\
$\indent \indent \indent \indent \indent  \indent  \indent x = 2 \times (cols \times dataSize)$ 

where $(m+1)$ and $cols$ are the sizes of rows and columns of the simplex tableau respectively. Thus, the size of each LP is $Y$ bytes, where $DType$ is the data type being used and $x$ being the size of array used for performing parallel reduction operation, the number $2$ in the equation $x = 2 \times (cols \times dataSize)$ signify the use of two auxiliary arrays. The size of the $cols$ is described in Subsection \ref{sub:The-Simplex-Algorithm}. Thus, if $S$ is the size of total global memory (in bytes) available in the GPU, then our threshold limit or the number of LPs that can be solved at a time is determined by the equation $N = \lfloor\frac{S}{Y}\rfloor$.
As the current limit on threads per block is $1024$ for GPU, thus, our implementation limits the size of an LP problem to 
 $511 \times 511$ for LP problems whose \emph{initial basic solution is feasible} and up to $340 \times 340$ for the class of LP problems with \emph{initial basic solution as infeasible}. This limit is defined by the inequality (\ref{eq:ThreadsPerBlock}) 
\begin{equation}
(var + slack + arti + 2) \le 1024 \label{eq:ThreadsPerBlock}
\end{equation}
 where $var$ is the number of variables (dimension of the LP problem), $slack$ is the number of slack variables (or constraints) and $arti$ is the number of artificial variables (if any) of the given LP as in the equation (\ref{eq:SingleLPSize}). 
This limitation can be overcome either by mapping a single thread to work on more than one data-instruction at a time or by mapping an LP problem with more than one thread blocks.

\subsection{Performance Analysis of Solving Batched LPs on GPU \label{sec:AnalyseMultiLPs-GPU}}
\begin{figure*}[!htb]
	\centering{}
	\subfloat[5-Dimension\label{fig:5dLP}]
	{\includegraphics[scale=0.4]{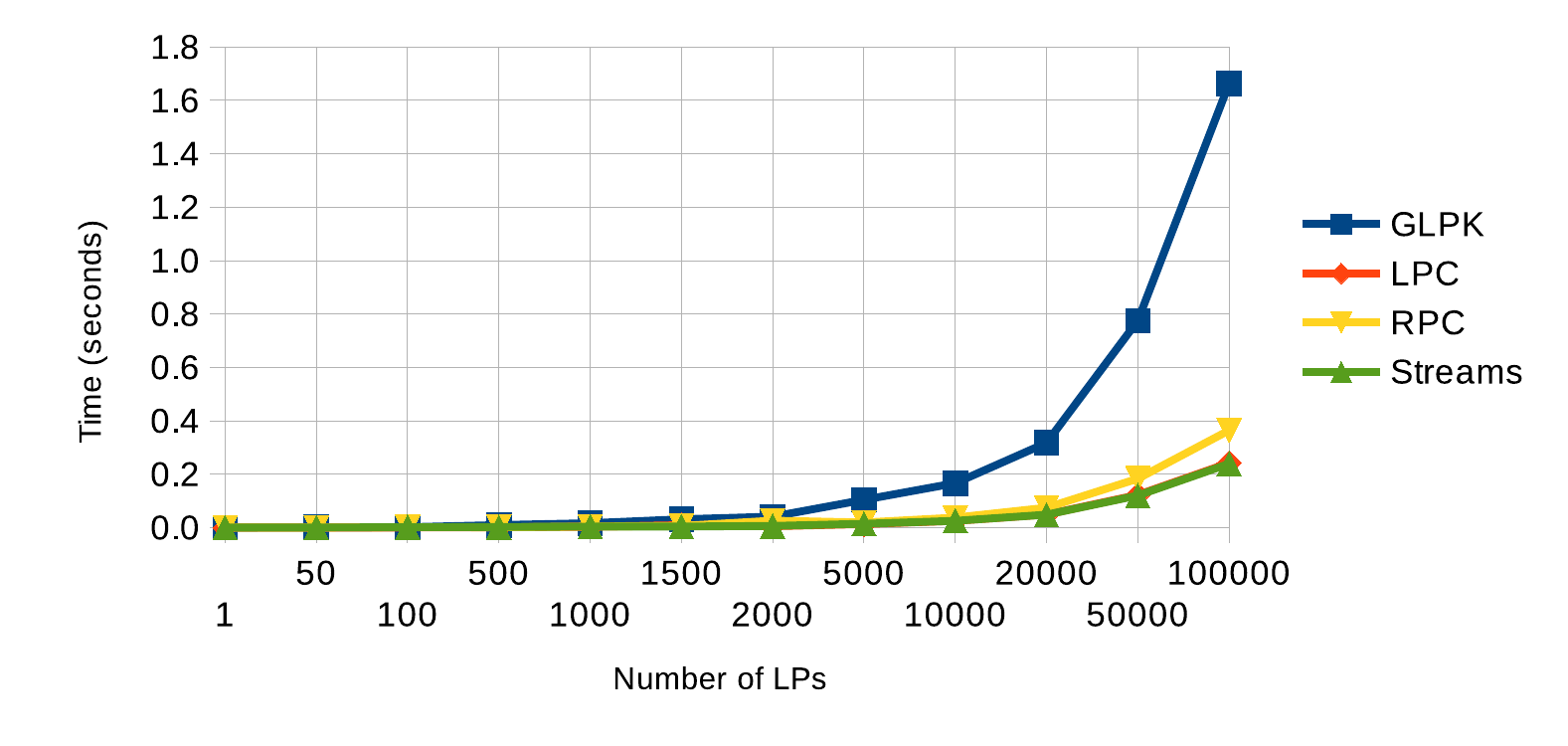}} 
	\subfloat[28-Dimension\label{fig:28dLP}]
	{\includegraphics[scale=0.4]{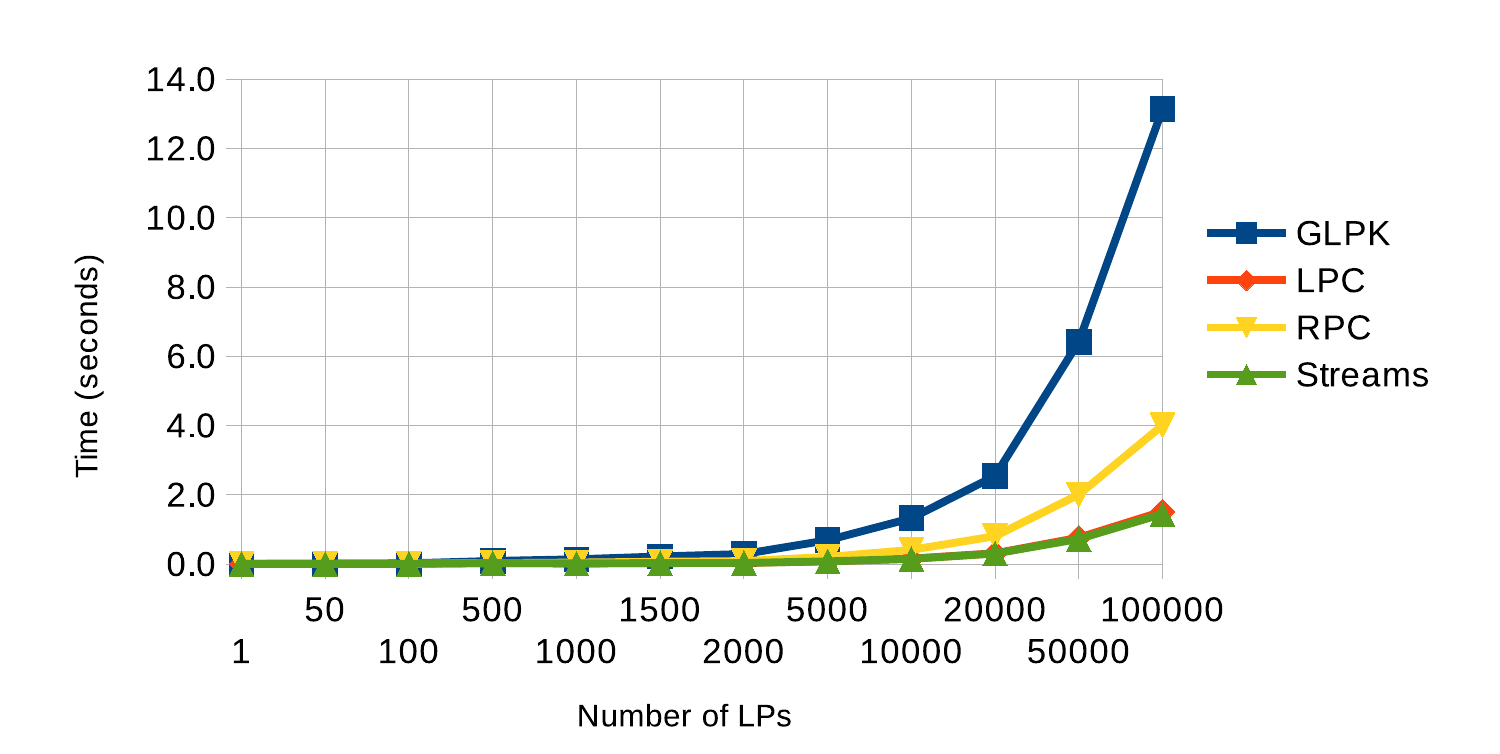}} \quad{}
	\subfloat[50-Dimension\label{fig:50dLP}]
	{\includegraphics[scale=0.4]{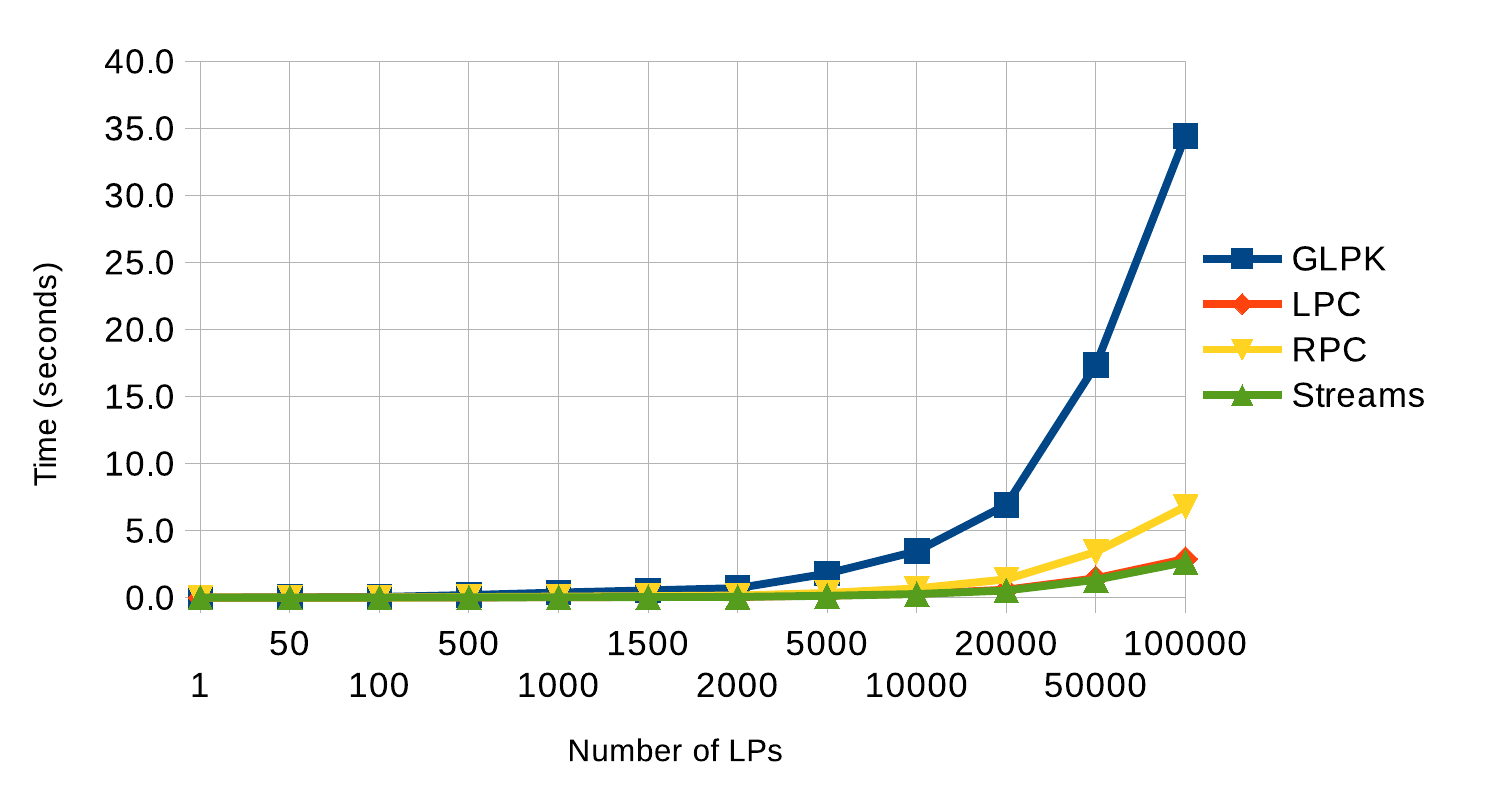}} 
	\subfloat[100-Dimension\label{fig:100dLP}]
	{\includegraphics[scale=0.4]{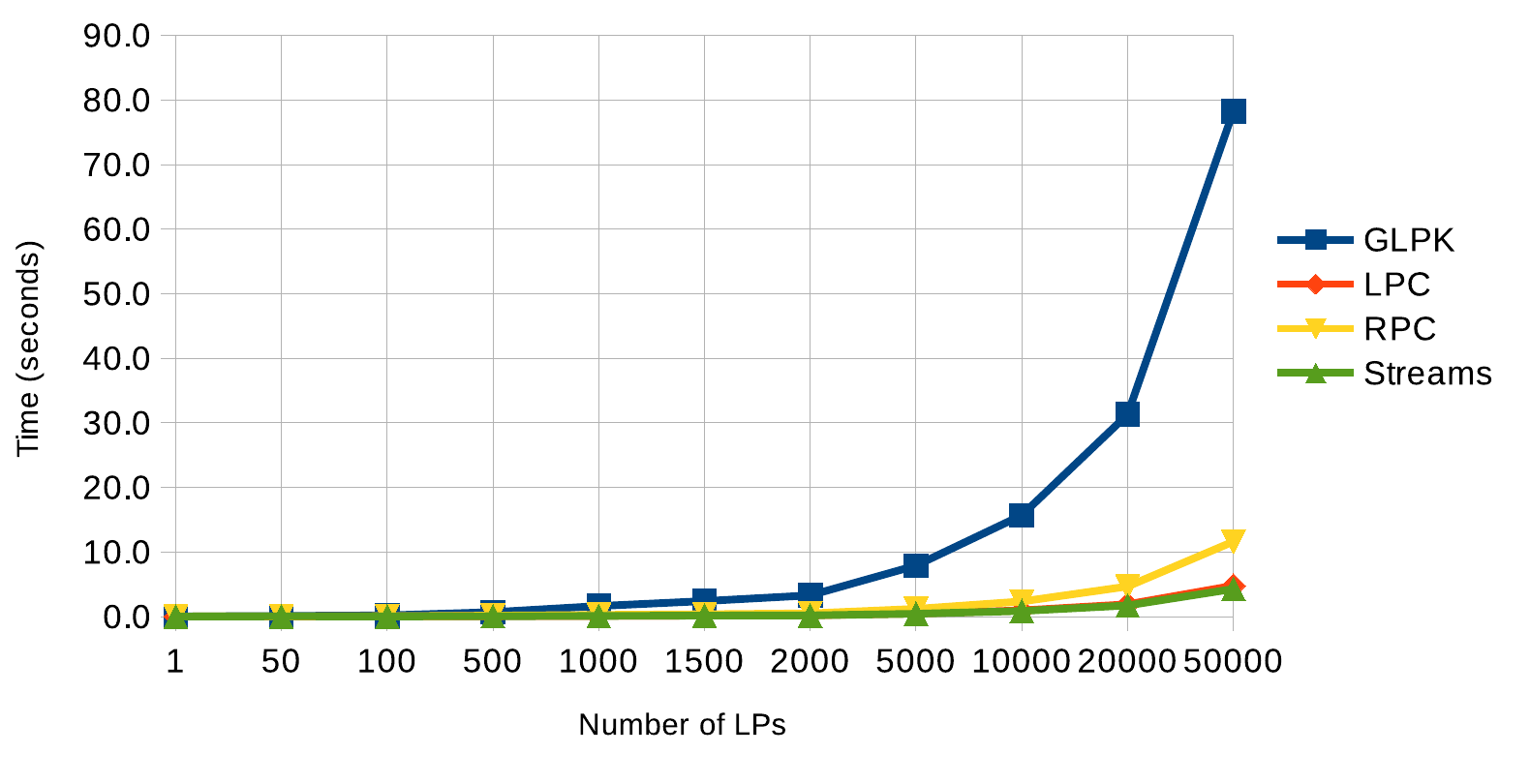}} \quad{}
	\subfloat[300-Dimension\label{fig:300dLP}]
	{\includegraphics[scale=0.4]{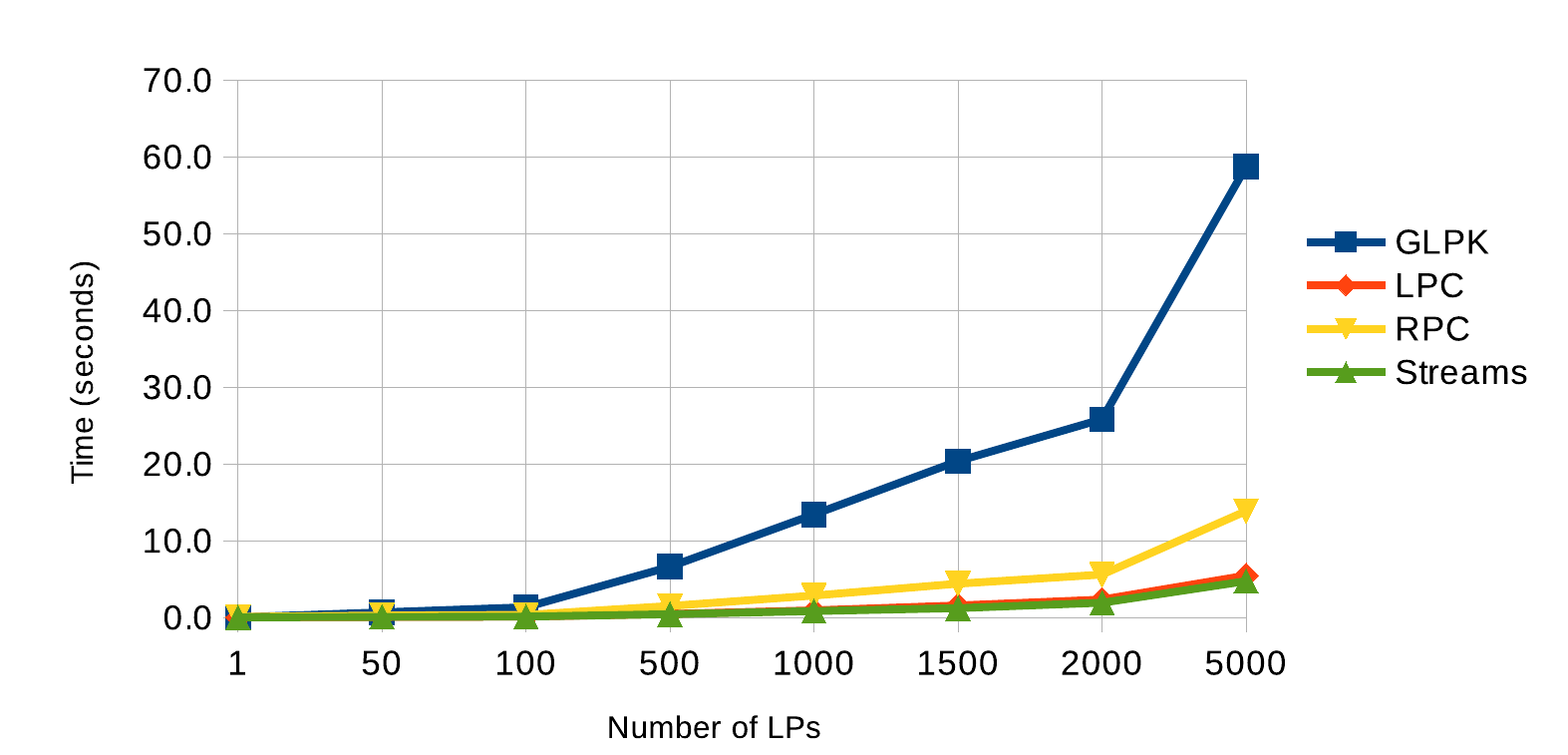}} 
	\subfloat[500-Dimension\label{fig:500dLP}]
	{\includegraphics[scale=0.4]{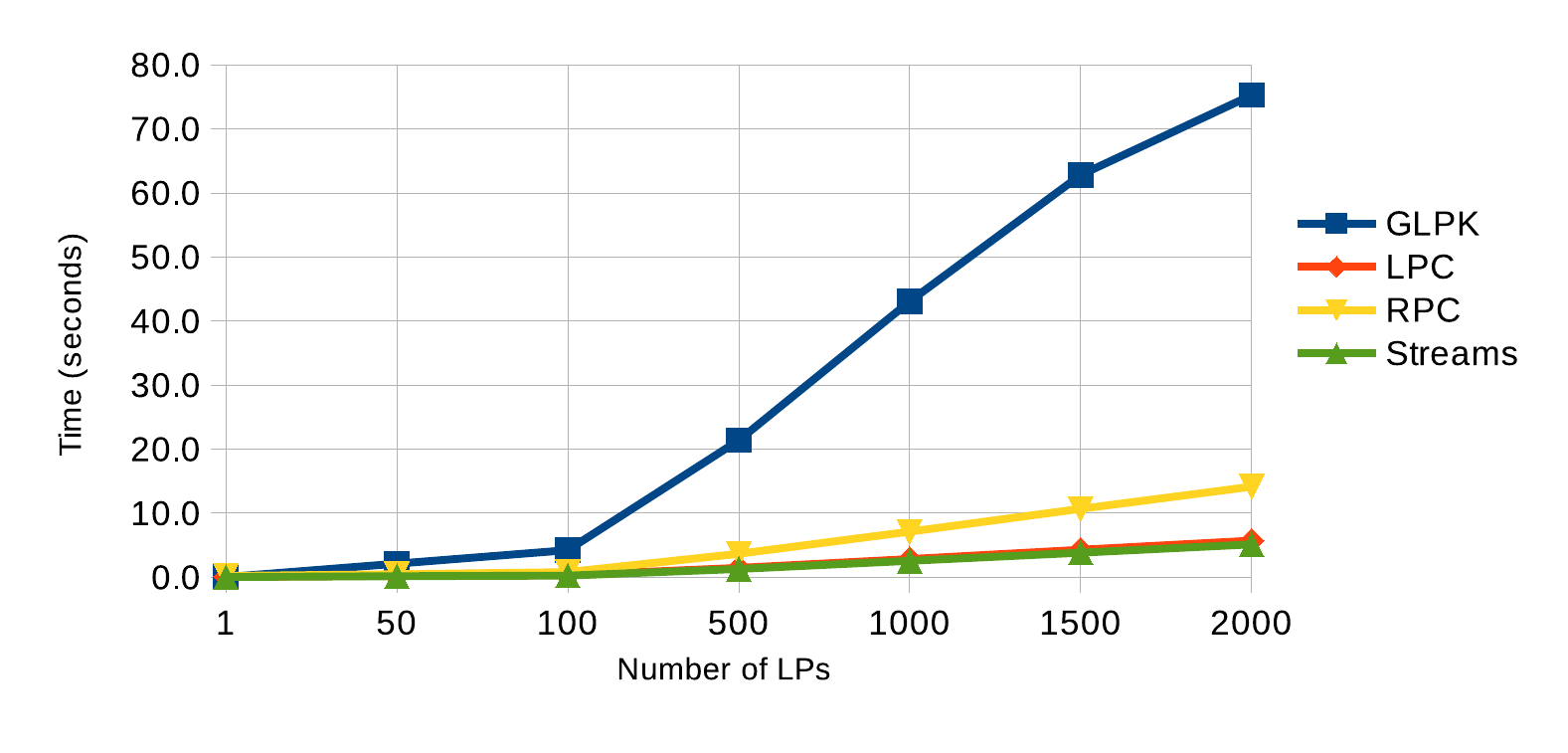}}
	
	\caption{Showing time taken to compute a batch of LPs for dimensions $5, 28, 50, 100, 300$ and $500$ respectively for the type of LPs with \textbf{initial basic solution as feasible}.
	\label{fig:TimeLPplotting}}
\end{figure*}

We performed our experiment in Intel Xeon E5-2670 v3 CPU, $2.30$GHz, $12$ Core (without hyper-threading), $62$GB RAM with Nvidia's Tesla K40c GPU card. The reported running time is an average over $10$ runs. We observed a maximum speedup of $16.43\times$ for 100-dimensional LP runs $20k$ LPs, using the LPC rule of pivot selection and a speedup of $6.74\times$ running $50k$ LPs of 100-dimension using the RPC rule of pivot selection, as compared to GLPK for LP problems which has \emph{initial basic solution as feasible}. A maximum speedup of $18.30\times$ is observed on $50k$ LPs of size 100 using streams with LPC rule, as compared to GLPK for LP problems which has \emph{initial basic solution as feasible}, as shown in Figure \ref{fig:TimeLPplotting}. We observed that for LPs of large size, our CUDA implementation performs better even with few LPs in parallel (e.g., batch-size$=50$ for 500 dimensional LP). However, for small size LPs, our CUDA implementation out-performs GLPK only for larger batch-size (e.g. $100$ and above for 5 dimensional LP). We also observe that the LPC pivot selection rule shows better performance than the RPC rule, although LPC involves the extra overhead of computing the maximum in each simplex iteration using parallel reduction. It is known that in most cases, the LPC rule converges to the optimum in less number of simplex iterations compared to the RPC rule. Therefore, we can deduce that the time taken in computing the extra iterations that are required using the RPC rule overshoots the performance gain by avoiding the maximum computation at each iteration.

\begin{figure*}[!htb]
\centering{}
   \includegraphics[width=\textwidth]{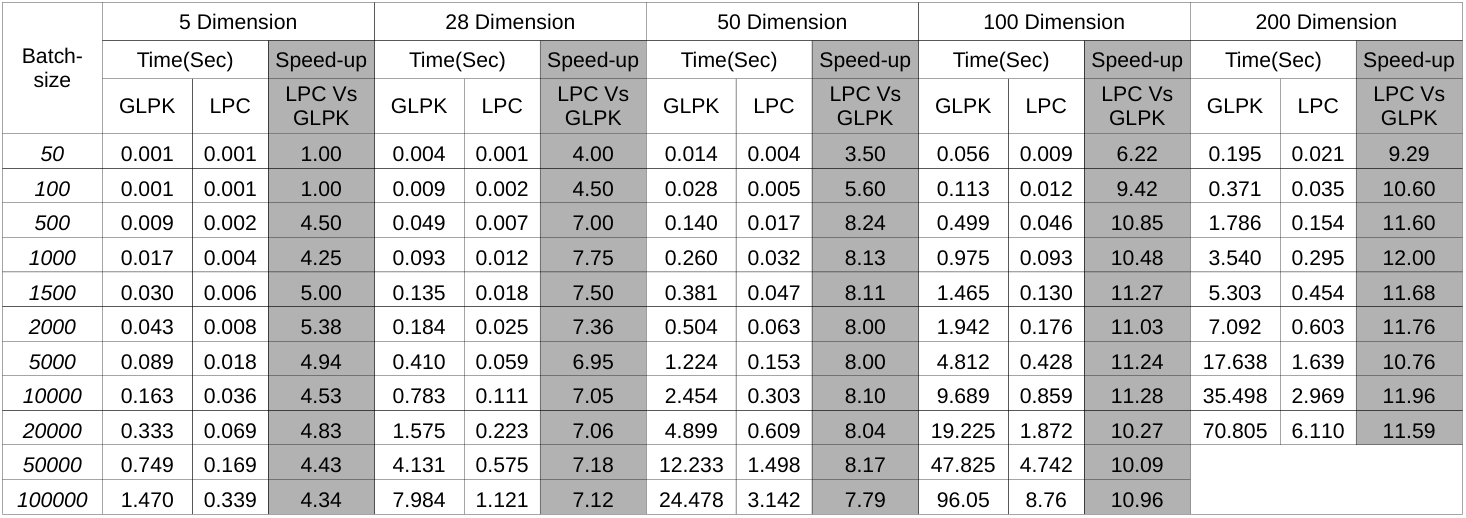}
\caption{Showing comparison between GLPK and GPU implementation for the type of LPs with \textbf{initial basic solution as infeasible}\label{tab:InfeasibleBasicSolution}}
\end{figure*}

For LP problems with \emph{infeasible initial basic solution}, though our implementation had to execute the kernel twice due to the two-phase simplex algorithm as described above in Section \ref{sec:LinearProgramming} (an extra overhead of data exchange between the two kernels), but we still observed that our implementation performed better than the GLPK library. We gain a maximum speedup of $11.96\times$ for $10k$ LPs of size $200$ using the LPC pivot selection rule compared to GLPK as shown in Figure \ref{tab:InfeasibleBasicSolution}.

While profiling the CUDA streams, we observed that for small sized LPs, the processing time of the kernel is much larger than the data transfer time as in Figure \ref{fig:profileResult} and so, the gain in performance of overlapping data transfer with kernel is also negligible as shown in \ref{fig:5dLP},\ref{fig:28dLP} and \ref{fig:50dLP}. But as the LP size increases (e.g., $500$) the size of data transfers are also significantly larger as in Figure \ref{fig:profileResult}. Hence, the operation of data transfer for all the streams (except the first) can be overlapped while the first kernel is in execution, thereby saving the time for data transfer in the rest of the stream. Thus, an extra $2-3\%$ gain in performance for LPs of larger dimensions is observed in our experiments, as shown in Figure \ref{fig:TimeLPplotting}, due to the overlapping of data transfer with the kernel's execution using the CUDA streaming technique.

\section{Implementation of Thread Safe GLPK \label{sec:parallelGLPK}}
GLPK (GNU Linear Programming Kit) is a commonly used open source and optimized linear programming solver library. However, the GLPK library is not thread safe, i.e., multiple threads running independent instances of GLPK objects may produce inconsistent results. In order to solve multiple LPs in parallel using GLPK, one may have a multi-process implementation which is not efficient since each process own its own memory space (when large numbers of GLPK objects are forked, large memory is going to be used). Experimentally, we observed that a multi-process GLPK implementation consumes double the size of memory as compared to its multi-threaded implementation. We identified that GLPK has a shared data which results in race condition on a multi-threaded setting. We modified the shared data (a pointer variable named \emph{tls} in the source file, ``glpenv02.c'') to thread-local in the source and we could ensure thread safety, enabling us to make parallel calls to GLPK in order to solve multiple LPs using multi-threading. Each independent thread, still solve a single LP sequentially. We use the OpenMP directives to create multiple execution threads, each making a call to an LP solver. 


\begin{figure*}[!htb]
	\centering{}
	\subfloat[10-Dimension\label{fig:5dLP}]
	{\includegraphics[width=4.9cm,height=3.5cm]{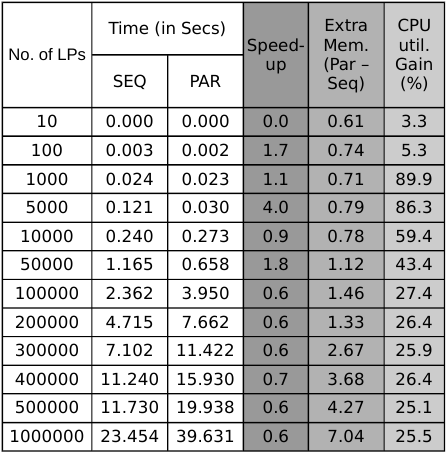}} \quad{}
	\subfloat[50-Dimension\label{fig:28dLP}]
	{\includegraphics[scale=1.0]{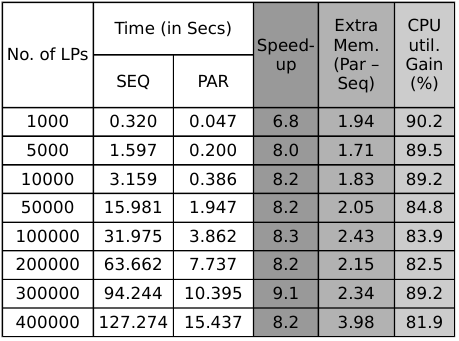}} \quad{}
	\subfloat[100-Dimension\label{fig:50dLP}]
	{\includegraphics[scale=1.0]{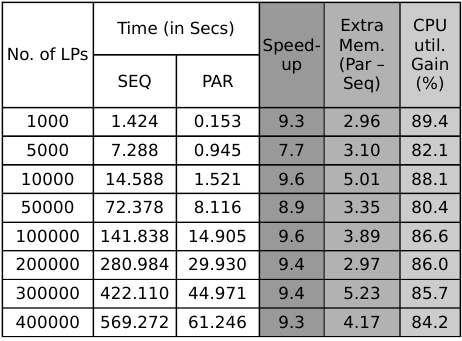}} \quad{}
	\subfloat[200-Dimension\label{fig:100dLP}]
	{\includegraphics[scale=1.0]{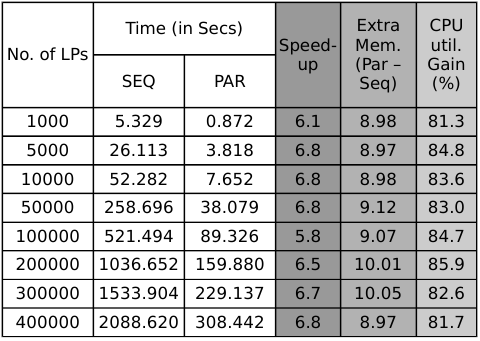}} \quad{}
	
	\caption{Showing comparison between multiple LPs solved in parallel using thread safe and sequential GLPK implementation on a 12-Core Intel Xeon processor\label{tab:SeqVsParGLPK}}
\end{figure*}

We performed multi-threading experiments with the thread safe GLPK on a $12$ cores Intel Xeon CPU E5-2670, 2.30GHz with 62.8 GB RAM, for an average of $10$ runs. 
 We observed a maximum of $9.6\times$ speedup for $1e5$ LPs of dimension 100, using the thread parallel GLPK as compared to sequential solving using GLPK. We have also recorded the overhead in memory (in Megabytes (MB)) incurred due to threading multiple GLPK objects in parallel as shown in Figure \ref{tab:SeqVsParGLPK} by the column labeled ``Extra mem. (Par - Seq)''. 
The table column ``CPU util. Gain (\%)'', is the difference of CPU utilization in parallel with that of the Sequential executions.
We observed that in our experimental setup, the thread parallel GLPK out performs the sequential GLPK only for LPs of dimension $16$ and above. For LPs with smaller dimensions, the penalty of thread creation and context switching is more than the the gain with parallelization.

Clearly, for large size LPs we observed the gain in speedup but at the cost of memory overheads due to threading multiple GLPK objects.

\section{CUDA Implementation of Special-Case LPs \label{sec:HyperboxLP}}
The feasible region of an LP given by its constraints defines a convex polytope. We observe that when the feasible region is a hyper-rectangle, which is a special case of a convex polytope, the LP can be solved cheaply. Equation (\ref{eq:HyperBoxEquation}) shows that maximizing the objective function is the sum of the results on $n$ dot products.

\begin{equation}\label{eq:HyperBoxEquation}
\maxi_{x \in \B} (\ell.x)=\sum_{i=1}^{n}\ell_{i}.h_{i},\text{where } h_{i} = 
\begin{cases}
a_i & \text{if $\ell_i<0$}\\
b_i & \text{otherwise}
\end{cases}
\end{equation}
where $\ell \in \Reals^n$ is the sampling directions over the given hyperbox $\B = \{x \in \Reals^n | x\in [a_1, b_1] \times ... \times [a_n, b_n] \}$.

The performance results of our GPU implementation of the hyperbox LP solver are presented in Table \ref{tab:BoundedLPinGPU}. In order to solve many LPs in parallel, we organize CUDA threads in a one-dimensional block of threads with each block used to solve an LP. Each block is made to consist of only 32 threads, the warp size. Within each block, we used only a single thread to perform the operations of the kernel. The operation $\sum_{i=1}^{n}l_{i}.h_{i}$, which can be performed using parallel reduction is expensive than computing sequentially, due to the overheads in implementing the parallel reduction technique. A preliminary introduction about this technique is introduced in the paper\cite{DBLP:conf/hvc/RayGDBBG15}.


\begin{table}[!htb]
\centering{}
\includegraphics[width=8.0cm,height=7.0cm]{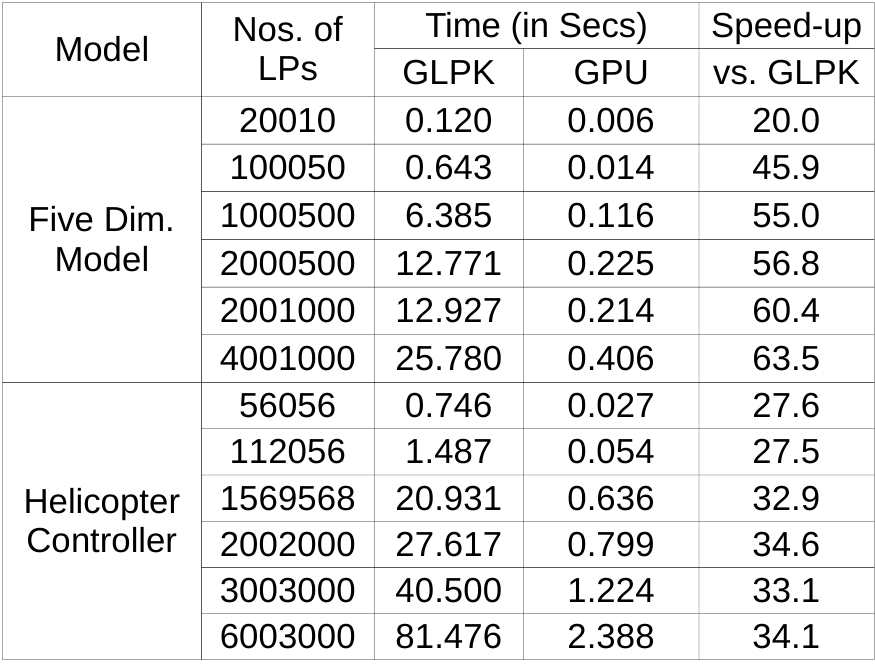}
\caption{Comparing GLPK with Hyperbox LP Solver in GPU}
\label{tab:BoundedLPinGPU}
\end{table}

We present the performance gained by our GPU implementation for hyperbox LPs in Table \ref{tab:BoundedLPinGPU}, as compared to solving  sequential using GLPK. In general, our hyperbox LP solver can simultaneously solve a batch of independent LPs (i.e. Each LP with a different set of constraints), but to keep the experimental setup same as in Table \ref{tab:supGPU} we consider the same LPs (i.e. All LPs with the same set of constraints) of five and 28 dimensions with large number of different objective functions. This setup enables an efficient sequential GLPK implementation 
The column labeled ``No. of LPs'' in Table \ref{tab:BoundedLPinGPU} indicate the total number of objective functions required to be solved for the same given LP problem.

We performed our experiment in Intel Q9950, 2.84Ghz, $4$ Core (no hyper-threading), 8GB RAM with GeForce GTX $670$ GPU card for an average of $10$ runs. We observed a $63.5\times$ speedup for $4001000$ LPs of 5-dimension and a $34.12\times$ speedup for $6003000$ LPs of 28-dimension, using our hyperbox LP solver in GPU as compared to the GLPK solver. 


\begin{table*}[!htb]
\centering{}
 \includegraphics[width=\textwidth]{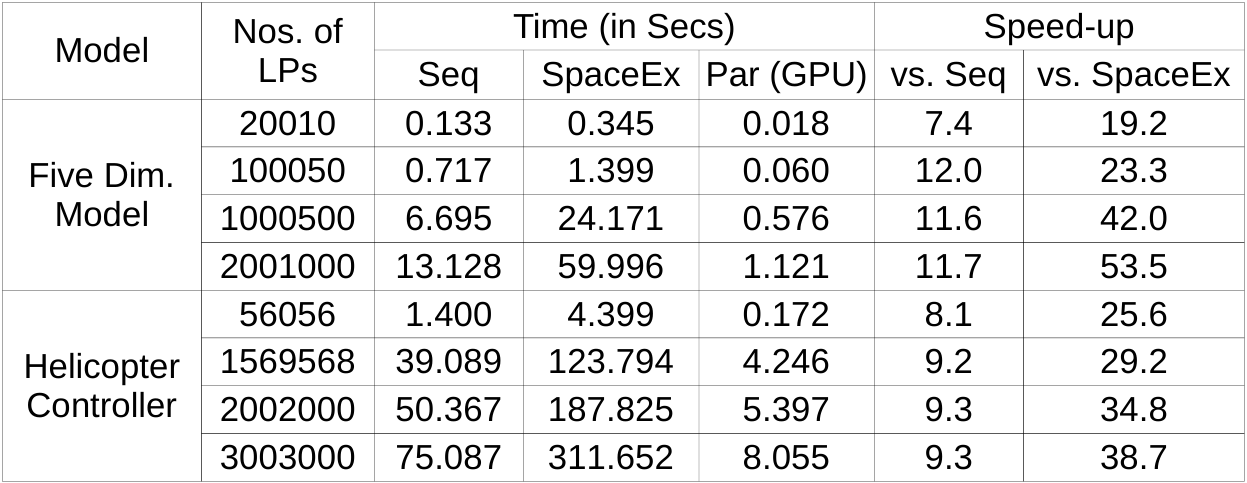}
\caption{Performance Speed-up in XSpeed using Hyperbox LP Solver}
\label{tab:supGPU}
\end{table*}

\section{Application of Parallel LP Solving in GPU \label{sec:HybridSystem}}
\begin{figure}[h]
\centering{}
\includegraphics[scale=0.6]{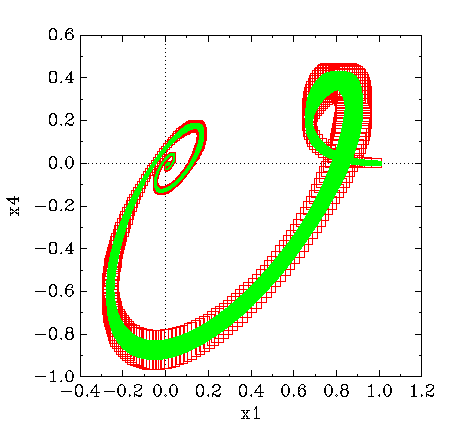}
\caption{Reachable State Space of a Five Dimensional Model with Different Precisions.}
\label{fig:reach}
\end{figure}

In the design of control systems, a standard technique of analysis is by mathematically modeling the control system design and computing the state-space of the model using exploration algorithms. Properties of the control system such as safety and stability can be analyzed with the computed state-space. In this section, we discuss two open-source tools that perform state-space exploration of linear systems with continuous dynamics. First is the tool SpaceEx \cite{FLGDCRLRGDM11} and the second is XSpeed \cite{DBLP:conf/hvc/RayGDBBG15}. These tools can analyze systems modeled as ordinary differential equations with uncertainty ($\dot{x} = A.x(t) + u$, $u \in \U$, $x \in \X_0$ at $t=0$), where the set $\U$ models the set of all possible control inputs. 
A conservative over-approximation of the exact reachable state space is computed by both these tools. A common state space computation algorithm in these tools compute the reachable state space as a union of convex sets, each having a symbolic representation in memory, known as the support function representation \cite{GirardLG08}. However, the algorithm requires to convert the convex sets from its support function representation to convex polytope representation, for certain operations to be efficient and for the visualization of the state space. Such a conversion to convex polytope representation loses precision since they provide an over-approximation of the original convex set. A support function of a convex set $\Omega \subset \Reals^d$ is a function that takes a vector $\ell \in \Reals^d$ as an argument and produces the real number $r = max_{x \in \Omega}\textit{ } \ell \cdot x$. The conversion from a support function representation to a polytope representation involves sampling the support function in a finite number of arguments $\ell$. The precision of the conversion depends on the number of function samples taken. It can be easily seen that sampling the support function of convex sets which are convex polytopes, is a linear programming problem. It is common in applications to have these convex sets $\Omega$ as convex polytopes. Therefore, these conversions results in many LPs to be solved. Figure \ref{fig:reach} shows the computed reachable state space of the model of a five dimensional system with different precisions, approximated by a union of convex polytopes using the tool XSpeed. It can be seen that the precision of the state space shown in green is better than the one shown in red. The former requires solving 1e6 LPs of size five whereas the later requires 1e5 LPs of same dimension to be solved. Therefore, we see that reachability analysis tools requires a large number of LP solving. We now briefly discuss two continuous systems of small dimension and illustrate the performance speedup obtained by using our library in the XSpeed tool with batched LP solving. 
    
 \subsection{Helicopter Controller}
It is a model of a twin-engined multi-purpose military helicopter with 8 continuous variables modeling the motion and 20 controller variables that govern the various controlling actions of the helicopter \cite{Skogestad:2005:MFC:1121635, FLGDCRLRGDM11}. We consider the initial input set $X_0$ to be an hyperbox and the non-deterministic input sets $\U$ as a point set. The fact that the given input set is an hyperbox enables us to perform all computation of support function samplings using our special-case LP solver in GPU as described earlier.

 \subsection{Five Dimensional System}
It is a model of a five dimensional linear continuous system as defined in \cite{Girard05}. We consider the initial input set $X_0$ as an hyperbox centered at (1,0,0,0) with sides of length 0.02 units. We take the non-deterministic input to be a point set as (0.01,0.01,0.01,0.01,0.01). We direct the reader to the paper \cite{Girard05} for details about the dynamics of the model. 

We performed our experiment in a system with Intel Q9950, 2.84Ghz, 4 Core (no hyper-threading), 8GB RAM with a GeForce GTX 670 card. The performance reported averages over 10 runs. In comparison to the sequential solving of LPs using the GLPK library call, we observed a maximum of $12\times$ and $9\times$ speedup with parallel LP solving in the GPU, for a five dimensional system and a helicopter controller benchmark respectively. When compared to the tool SpaceEx, we observed a maximum of $54\times$ and $39\times$ speedup in XSpeed using our CUDA implementation, for a five dimensional system and a helicopter controller model respectively, as shown in Table \ref{tab:supGPU} (a section of this result has also been published in the paper \cite{DBLP:conf/hvc/RayGDBBG15}).

\section{Conclusion \label{sec:Conclusion}}
We present a CUDA implementation to solve multiple LPs of small to medium size simultaneously in a GPU. We explain the implementation choices to have a coalescent memory access, efficient load balancing and efficient CPU-GPU memory copy operation using CUDA streams. We show the techniques that have been used to implement the simplex algorithm, like parallel reductions and loop interchange. We experiment with two pivot selection rules in the simplex algorithm to observe its performance in GPU. We deduce with experiments that the LPC rule shows better performance than the RPC rule in GPU even though LPC involves the extra overhead of parallel reduction. We present a thread safe GLPK implementation and its experimental evaluation in solving many LPs with multi-threading in a multi-core CPU. We demonstrate significant performance speedup with the multi-threaded implementation. It is observed that LPs can be solved very cheaply when its feasible region that given by its constraints is a hyperbox, i.e., a subclass of general convex polytopes. We implement the solution of such special case LPs in CUDA and report significant performance improvement when compared to solving using GLPK. Lastly, We illustrate an application which involves many LP solving of small to moderate size and we show the performance speedup of the application with our batched LP solver library.

\section*{Acknowledgments}
This work was supported by the National Institute of Technology Meghalaya, India and by the the DST-SERB, GoI under project grant No. YSS/2014/000623.


\bibliography{mybibfile}

\end{document}